\documentclass{revtex4}
\usepackage{graphicx}
\usepackage{fancyhdr}
\usepackage{amsmath}
\usepackage{color}

\begin{document}

\title{Overview of Electroweak Baryogenesis}

%

\author{Eibun Senaha}
\affiliation{School of Physics, KIAS, Seoul 130-722, Korea}
\affiliation{Department of Physics, Nagoya University, Nagoya 464-8602, Japan}

\begin{abstract}
Electroweak baryogenesis is briefly overviewed with emphasis on nondecoupling loop properties
of the scalar particles. In particular, quantum corrections to the triple Higgs boson coupling 
is discussed as a probe of electroweak baryogenesis.
\end{abstract}

\maketitle

\thispagestyle{fancy}


\section{Introduction}
According to the cosmological data, the baryon asymmetry of the universe (BAU) is found to be~\cite{Beringer:1900zz}
\begin{align}
\eta_{\rm CMB}  &= \frac{n_b}{n_\gamma} = (6.23\pm0.17)\times 10^{-10}, \\
\eta_{\rm BBN}  &= \frac{n_b}{n_\gamma} = (5.1-6.5)\times 10^{-10},~(95\%~{\rm C.L.}).
\end{align}
Clarification of the origin of the BAU is one of the greatest challenges in particle physics and cosmology.
If the BAU is generated before $T=\mathcal{O}(1)$ MeV, the light element abundances
(D,${}^3{\rm He}$,${}^4{\rm He}$,${}^7{\rm Li}$) can be explained by the
standard Big-Bang cosmology. To get the right $\eta$ (which is called baryogenesis) from an initially
baryon symmetric universe, the so-called 
Sakharov's conditions should be satisfied~\cite{Sakharov:1967dj}: 
(i) baryon number $(B)$ violation, (ii) C and CP violation, (iii) departure from thermal equilibrium.
To this end, a lot of scenarios have been proposed so far~\cite{Shaposhnikov:2009zzb}. 
From an experimental point of view, 
electroweak baryogenesis (EWBG)~\cite{ewbg} among others is the most testable scenario,
and thus it is in urgent need of detailed analysis in the LHC and future ILC eras.
Since it is not very trivial to satisfy all the Sakharov's conditions within the EWBG framework, 
some distinctive features are required in the feasible region  
and thus may yield striking signals to the low energy observables.

In this talk, the EWBG is reviewed focusing especially on nondecoupling loop properties of scalar particles
that may help for realizing the strong first-order phase transition (EWPT).
As a consequence of the nondecoupling scalar loop, 
quantum corrections to the triple Higgs boson coupling is discussed.

\section{EWBG mechanism}
Foregoing Sakharov's conditions in the EWBG are satisfied as follows:
(i) $B$ violation is realized by an anomalous process at finite temperature 
(conventionally referred as {\it sphaleron process} although the sphaleron solution does not exist
in the symmetric phase.)
(ii) C is maximally violated by the chiral gauge interactions, 
CP violation comes from the Cabbibo-Kobayashi-Maskawa matrix~\cite{CKM} or other complex parameters 
once the standard model (SM) is extended.
(iii) Out of equilibrium is realized by the first-order EWPT with bubble nucleation and expansion.

\begin{figure}[t]
\centering
\includegraphics[width=5cm]{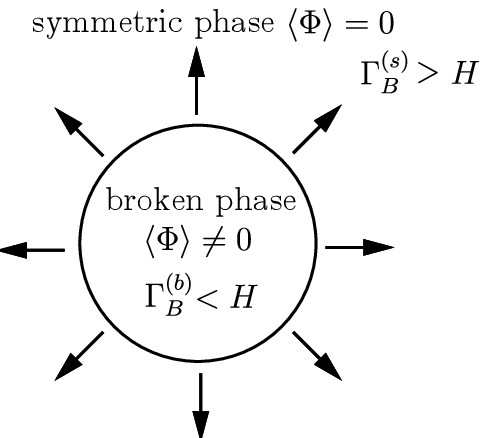}\hspace{1cm}
\includegraphics[width=6cm]{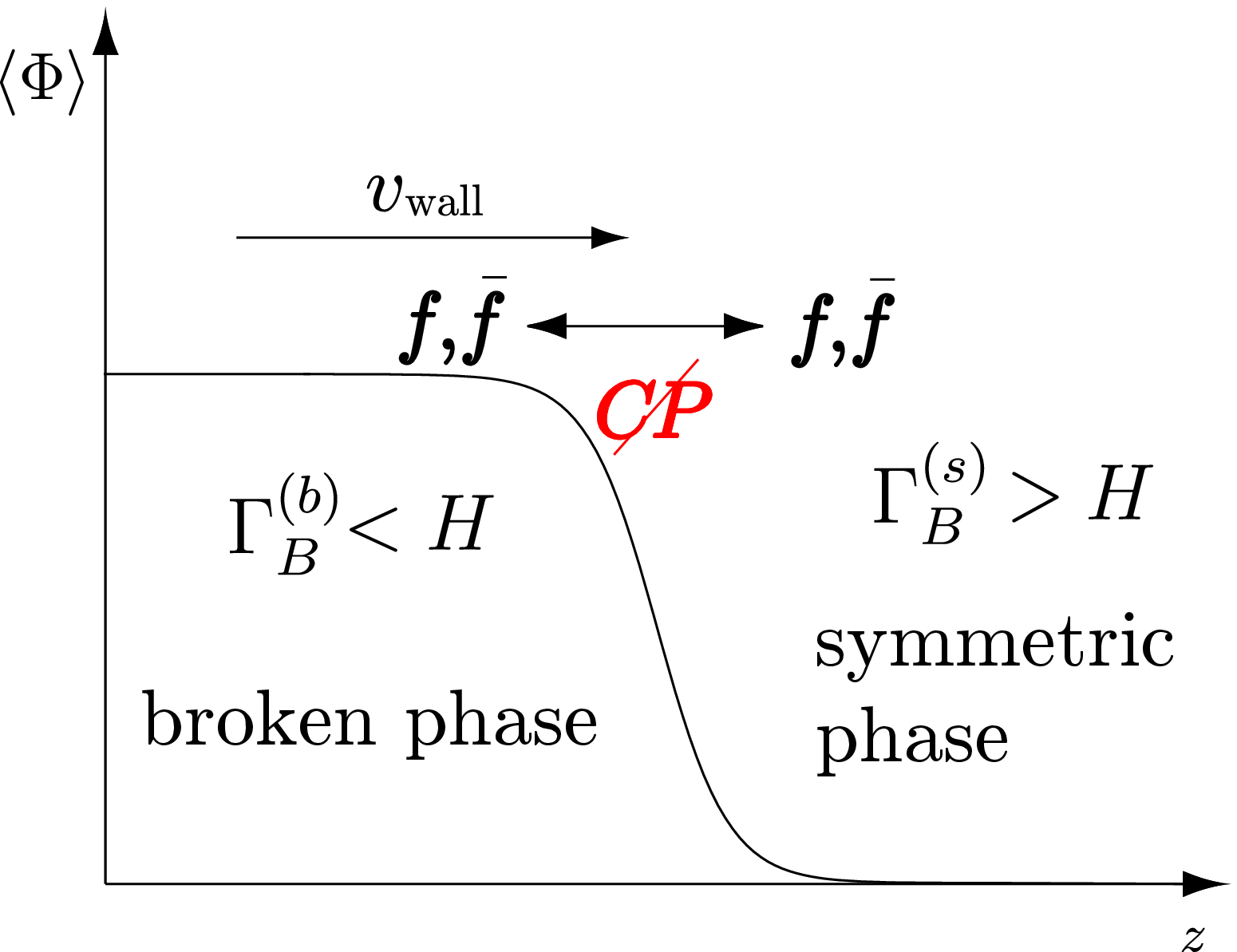}
\caption{(Left) The bubble expansion. (Right) The Higgs VEV as function of $z$ which is the direction of the bubble expansion.} 
\label{fig:bubble}
\end{figure}

The left panel of Fig.~\ref{fig:bubble} shows a schematic picture of the expanding bubble wall.
The inside of the bubble corresponds to the broken phase where
the Higgs vacuum expectation value (VEV) is nonzero, $\langle\Phi\rangle\neq0$
while the outside of it represents the symmetric phase where $\langle\Phi\rangle=0$.
In the right panel of Fig.~\ref{fig:bubble}, $\langle\Phi\rangle$ is depicted as a function of $z$ 
which is a direction of the bubble expansion.
The outline of the EWBG is as follows.
\begin{itemize}
\item[1.] Because of $CP$ violation induced by interactions between the bubble 
and the particles in the plasma, 
chiral charges are asymmetrized.
\item[2.] They diffuse into the symmetric phase and accumulate.
\item[3.] $B$ is generated via sphaleron process. 
\item[4.] After decoupling of the sphaleron process in the broken phase, $B$ is fixed. 
\end{itemize}
The last step may leave a detectable footprint in low energy observables.
In the following, we will look into one of such possibilities.

\section{Sphaleron decoupling condition}
In order to preserve the generated BAU via the sphaleron process in the symmetric phase, 
the $B$-changing rate in the broken phase ($\Gamma_B^{(b)}$) must be sufficiently suppressed.
Namely, 
\begin{align}
\Gamma_B^{(b)}(T)\simeq ({\rm prefactor})e^{-E_{\rm sph}(T)/T} 
< H(T)\simeq 1.66\sqrt{g_*(T)}T^2/m_{\rm P} \label{GamvsH}
\end{align}
is satisfied, where $E_{\rm sph}$ denotes the sphaleron energy, 
$g_*$ is the degrees of freedom of relativistic particles in the plasma
($g_*=106.75$ in the SM) and $m_{\rm P}$ stands for the Planck mass 
which is about $1.22\times 10^{19}$ GeV.
Since $E_{\rm sph}$ is proportional to the Higgs VEV (denoted by $v$), 
Eq.~(\ref{GamvsH}) can be realized if the EWPT is strongly first-order.
Conventionally, the sphaleron energy is parametrized as $E_{\rm sph}(T)=4\pi v(T)\mathcal{E}(T)/g_2$, 
where $g_2$ denotes the SU(2) gauge coupling constant. Eq.~(\ref{GamvsH}) is then cast into the form
\begin{align}
\frac{v(T)}{T} > \frac{g_2}{4\pi \mathcal{E}(T)}
\Big[
	42.97+\mbox{log corrections}
\Big].\label{sph_dec}
\end{align}
The dominant contributions on the right-hand side is $\mathcal{E}(T)$
while the log corrections that mostly come from the zero mode factors of the fluctuations 
about the sphaleron typically amount to about 10\%~\cite{Funakubo:2009eg}.

As an illustration, we evaluate the sphaleron energy at zero temperature, $\mathcal{E}(0)$, 
within the SM~\cite{sph_SM}.
Since the ${\rm U(1)}_Y$ contribution is sufficiently small~\cite{sph_wU1Y}, 
it is enough to confine ourself to the ${\rm SU(2)}_L$ gauge-Higgs system.
To find the sphaleron solution, we adopt a spherically symmetric configurations ansatz
with a noncontractible loop~\cite{sph_SM}. 

\begin{figure}[t]
\centering
\includegraphics[width=5cm]{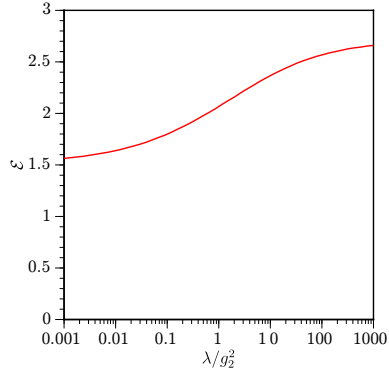}
\caption{The dimensionless sphaleron energy $\mathcal{E}(0)$ vs. $\lambda/g_2^2$.} 
\label{fig:rEsph_lamg2}
\end{figure}

In Fig.~\ref{fig:rEsph_lamg2}, $\mathcal{E}(0)$ is plotted as a function of $\lambda/g_2^2$.
We can see that as $\lambda/g_2^2$ increases, $\mathcal{E}$ increases. 
For the Higgs boson with a mass of 126 GeV, which corresponds to $\lambda\simeq 0.13$, 
one finds $\mathcal{E}(0)\simeq1.92$.
With this value, Eq.~(\ref{sph_dec}) becomes
\begin{align}
\frac{v(T)}{T}>1.16,
\end{align}
where only the dominant contributions are retained on the right-hand side in Eq.~(\ref{sph_dec}).
Note that the use of $\mathcal{E}(0)$ in the decoupling criterion leads to somewhat underestimated results
since $\mathcal{E}(T)<\mathcal{E}(0)$.
In the MSSM, using the finite-temperature effective potential at the one-loop level, 
$v(T_N)/T_N>1.38$ is obtained, where the sphaleron energy as well as the translational 
and rotational zero mode factors of the fluctuation around the sphaleron 
are evaluated at a nucleation temperature ($T_N$) which is somewhat below 
$T_C$~\cite{Funakubo:2009eg}.

From the argument here, the degree to which the first-order EWPT is strengthen
is one of the central questions to be answered for successful baryogenesis.
There are several ways to achieve the strong first-order EWPT (see a recent study~\cite{Chung:2012vg}).
In what follows, we will focus on the thermally driven case, specifically.

Since most of collider probes of the EWBG are intimately related to the sphaleron decoupling condition,
the reduction of theoretical uncertainties in this condition is indispensable for reaching 
the definitive conclusion of viability of the EWBG. 
So far, a high-temperature expansion has been exclusively used to 
evaluate the sunset diagrams in the two-loop analysis of the EWPT.
Recently, the validity of such a high-temperature expansion
is investigated in the Abelian-Higgs and its extended models in~\cite{Funakubo:2012qc}. 
For gauge dependence issues in the perturbative analysis of the EWPT, see e.g. 
\cite{Patel:2011th,Wainwright:2012zn,Garny:2012cg} and references therein.

\section{Electroweak phase transition}
Here, we explicitly demonstrate why the SM EWBG fails, 
which may give a signpost searching for new physics.  
Using a high-temperature expansion, the one-loop effective potential at finite temperature is reduced to
\begin{equation}
V_{\rm eff}(\varphi; T)\simeq D(T^2-T^2_0)\varphi^2
-ET\varphi^3+\frac{\lambda_{T}}{4}\varphi^4,
\label{Veff-hte}
\end{equation}
where
\begin{align}
D &= \frac{1}{8v^2}\Big(2 m_W^2+m_Z^2+2m_t^2\Big),\quad
E = \frac{1}{4\pi v^3} \Big(2m_W^3+m_Z^3\Big)\simeq 10^{-2},\\
\lambda_T &= 
 \frac{m_h^2}{2 v^2} 
 \bigg[ 1 
- \frac{3}{8\pi^2 v^2 m_h^2} 
 \bigg\{ 
 2 m_W^4\log\frac{m_W^2}{\alpha_BT^2}  
+ m_Z^4\log \frac{m_Z^2}{\alpha_BT^2}  
- 4 m_t^4\log \frac{m_t^2}{\alpha_FT^2}\bigg\}\bigg],
\end{align}
with $\log\alpha_B=2\log4\pi-2\gamma_E^{}\simeq3.91$ and
$\log\alpha_F=2\log\pi-2\gamma_E^{}\simeq1.14$.
Appearance of the cubic term with the negative coefficient dictates that
the EWPT should be first-order.
Note that since the origin of the cubic term is the zero Matsubara frequency mode,
the only bosonic thermal loops contribute to $E$. 

The critical temperature $T_C$ is defined by a temperature at which $V_{\rm eff}(\varphi; T)$
has two degenerate minima. At $T_C$, $V_{\rm eff}$ takes the form
\begin{equation}
V_{\rm eff}(\varphi; T_C)=\frac{\lambda_{T_C}}{4}\varphi^2(\varphi-v_C)^2,\quad
v_C=\frac{2ET_C}{\lambda_{T_C}}.
\end{equation}

As we discussed in the previous section, $v_C/T_C\gtrsim 1$ should hold
to avoid the washout by the sphaleron in the broken phase.
Since $\lambda_{T_C}\simeq m_h^2/2v^2$, one obtains the upper bound of the Higgs boson mass as
\begin{equation}
m_h\lesssim 2v\sqrt{E} \simeq 48~\mbox{GeV}.
\end{equation}
This mass range has been already excluded by the LEP experiments.
According to nonperturbative studies, the EWPT in the SM is a crossover for $m_h\gtrsim 73$ GeV~\cite{sm_ewpt}.

From the above argument, one of the straightforward way-outs is to enhance $E$ 
by adding the bosonic degrees of freedom.
However, in the case of scalar particle, the degree to which $E$ is enhanced highly depends 
on the loop properties, namely, {\it decoupling} or {\it nondecoupling}.

Suppose a mass of a scalar is given by
\begin{align}
m^2 = M^2+\tilde{\lambda} \varphi^2,
\end{align}
where $M$ is a mass parameter in the Lagrangian or a thermally corrected mass at finite temperature,
and $\tilde{\lambda}$ is a coupling constant between the Higgs boson and the scalar.
Depending on the magnitudes of $M$ and $\tilde{\lambda}$, two cases are conceivable:
\begin{align}
V_{\rm eff} \ni
\left\{
	\begin{array}{l}
	-\tilde{\lambda}^{3/2}T\varphi^3
\left(
	1+\frac{M^2}{\tilde{\lambda}\varphi^2}
\right)^{3/2},\quad \mbox{for $M^2\ll\tilde{\lambda}\varphi^2$}, \\
-|M|^3T
\left(
	1+\frac{\tilde{\lambda}\varphi^2}{M^2}
\right)^{3/2}, \quad \mbox{for $M^2\gtrsim\tilde{\lambda}\varphi^2$}.
	\end{array}
\right.\label{Veff_2limits}
\end{align}
Note that the first case (what we call nondecoupling~\footnote{Of course, the effect of the scalar loop would be eventually  Boltzmann suppressed in the limit of $m\gg T$.}) can generate the cubic-like term so the $E$ 
is enhanced
while the latter (decoupling) does not. 
This is the reason why the Higgs boson whose mass is given by $m_h^2(\varphi)=-m_h^2/2+3\lambda \varphi^2$
does not contribute to $E$ significantly. 

In summary, what one needs for realizing the strong first-order EWPT via 
the thermally driven mechanism are (1) relatively large coupling constant $\tilde{\lambda}$ 
and (2) small mass parameter $M$
\footnote{In this regard, scale invariant theories might be preferable. See, for example, a study of the EWPT
in the massless 2HDM~\cite{Funakubo:1993jg}.}.
Phenomenological implications of such a nondecoupling scalar
are discussed below.

\section{Triple Higgs boson coupling as a probe of EWBG}
As an interesting consequence of the strong first-order EWPT, 
we consider quantum corrections to the triple Higgs boson coupling.
Here, we take the two Higgs doublet model (2HDM) as an illustration~\cite{Kanemura:2004ch}.
We impose a $Z_2$ symmetry to avoid Higgs-mediated flavor changing neutral current processes
at the tree level. In the Higgs potential, however, such a discrete symmetry is softly broken
by $\Phi_1^\dagger\Phi_2$ term. Specifically, 
the Higgs potential is given by
\begin{align}
V_{\rm 2HDM}
&= m_1^2\Phi_1^\dagger\Phi_1+m_2^2\Phi_2^\dagger\Phi_2
	-(m_3^2\Phi_1^\dagger\Phi_2+{\rm h.c.}) \nonumber\\
&\quad +\frac{\lambda_1}{2}(\Phi_1^\dagger\Phi_1)^2
	+\frac{\lambda_2}{2}(\Phi_2^\dagger\Phi_2)^2
	+\frac{\lambda_3}{2}(\Phi_1^\dagger\Phi_1)(\Phi_2^\dagger\Phi_2)
	+\frac{\lambda_4}{2}(\Phi_1^\dagger\Phi_2)(\Phi_2^\dagger\Phi_1)
	+\left[\frac{\lambda_5}{2}(\Phi_1^\dagger\Phi_2)^2+{\rm h.c.}\right].
\end{align}
In what follows, we assume that CP is not broken for simplicity.
Among the 8 model parameters, $v=\sqrt{v_1^2+v_2^2}$ ($v_{1,2}$ are VEVs of $\Phi_{1,2}$) 
and $m_h$  are known, so the free parameters are:
\begin{align}
m_H, \quad m_A,\quad m_{H^\pm},\quad \alpha,\quad \tan\beta=\frac{v_2}{v_1}, 
\quad M^2\equiv \frac{m_3^2}{\sin\beta\cos\beta},
\end{align}
where the first 3 $m$'s denote the masses of the CP-even Higgs boson, 
the CP-odd Higgs boson and the charged Higgs bosons.  
$\alpha$ is defined as the mixing angle between $h$ and $H$. 
The normalization $1/\sin\beta\cos\beta$ in $M^2$ 
is chosen in such a way that $M$ would be a CP-odd Higgs boson mass in the MSSM-like limit.

In this model, the dominant one-loop corrections to the triple Higgs boson coupling ($\lambda_{hhh}$) 
in the SM-like limit $(\sin(\beta-\alpha)=1)$ is found to be~\cite{HHH_2HDM}
\begin{align}
\lambda_{hhh}^{\rm 2HDM}
\simeq \frac{3m_h^2}{v}
\left[
	1+\sum_{\Phi=H, A, H^\pm}\frac{c}{12\pi^2}\frac{m_\Phi^4}{m_h^2v^2}\left(1-\frac{M^2}{m_\Phi^2}\right)^3
\right],
\end{align}
where $c=1$ for $\Phi=H, A$ and $c=2$ for $\Phi=H^\pm$ and the masses of the heavy Higgs bosons ($m_\Phi$)
have the form
\begin{align}
m_\Phi^2 = M^2+\lambda_iv^2,
\end{align}
with $\lambda_i$ being some combinations of the Higgs quartic couplings.
Since $m_\Phi^2$ is composed of the two parts, we may consider the two limits:
$M^2\ll \lambda_iv^2$ and $M^2\gtrsim \lambda_i v^2$. Correspondingly, 
the dependences of the heavy Higgs boson loop effects on $\lambda_{huh}^{\rm 2HDM}$ 
are reduced to
\begin{align}
\delta\lambda_{hhh}^{\Phi} \sim
\left\{
	\begin{array}{l}
	m_\Phi^4,\quad \mbox{for $M^2\ll\lambda_iv^2$}, \\
	\frac{1}{m_\Phi^2}, \quad \mbox{for $M^2\gtrsim\lambda v^2$}.
	\end{array}
\right.\label{delhhh_2limits}
\end{align}
The former corresponds to the nondecoupling limit while the latter is the ordinary decoupling limit
in which the effect would vanish in the large mass limit.
So far, we implicitly assume $M^2\ge 0$. 
The possibility of $M^2<0$ may be interesting~\footnote{It is pointed out by Abdesslam Arhrib.}.

\begin{figure}[t]
\centering
\includegraphics[width=8cm]{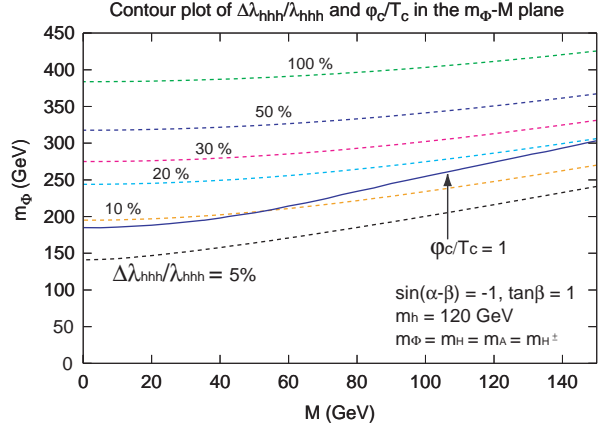}
\caption{The correlation between $\Delta\lambda_{hhh}^{\rm 2HDM}/\lambda_{hhh}^{\rm SM}$ and $v_C/T_C$.} 
\label{fig:Delhhh_vs_vcTc}
\end{figure}

Now we are in a position to discuss a relationship between strength of the first-order EWPT
and the loop effects on $\lambda_{hhh}$. Clearly, there is the strong correlation between them
as can be seen from Eqs.~(\ref{Veff_2limits}) and (\ref{delhhh_2limits}).
In Fig.~\ref{fig:Delhhh_vs_vcTc}, we quantify this correlation. 
$v_C(=\sqrt{v_1^2(T_C)+v_2^2(T_C)})$ and $T_C$ are determined 
by the numerical evaluation of the ring-improved one-loop effective potential.
We take $\sin(\beta-\alpha)=\tan\beta=1$, 
and assume that all the heavy Higgs boson masses are degenerate.
At the time of this study back in 2004, the Higgs boson mass was unknown, 
and thus it was set to 120 GeV. 
We expect that $m_h=126$ GeV case would not change significantly.
The deviation of $\lambda_{hhh}^{\rm 2HDM}$
from its SM value is defined by $\Delta\lambda_{hhh}^{\rm 2HDM}/\lambda_{hhh}^{\rm SM}$, where
$\Delta\lambda_{hhh}^{\rm 2HDM} = \lambda_{hhh}^{\rm 2HDM}-\lambda_{hhh}^{\rm SM}.$
We can see that for $v_C/T_C>1$ to be fulfilled, the heavy Higgs boson should be greater than about 200 GeV.
In such a case, $\Delta\lambda_{hhh}^{\rm 2HDM}/\lambda_{hhh}^{\rm SM}$ would be larger than
10\%. Remarkably, if the heavy Higgs bosons have the strong nondecoupling loop effects, 
$\Delta\lambda_{hhh}^{\rm 2HDM}/\lambda_{hhh}^{\rm SM}$ can reach 100\%.
Note that it does not necessarily imply the breakdown of perturbation since the Higgs quartic couplings
appearing in the tree-level $\lambda_{hhh}$ and those in the one-loop expression are essentially different
in a sense that the combinations of $\lambda_{1-5}$ are different.
Actually, $\lambda_{1-5}<4\pi$ are still maintained.

Let us consider the MSSM case next.
The EWBG scenario in this model is the so-called ``light stop scenario (LSS)~\cite{LSS}",
which is now on the verge of being excluded. (see e.g. \cite{LHCtension,Carena:2012np}).

In order to realize the physical Higgs boson mass, the left-handed stop SUSY breaking mass ($m_{\tilde{q}}$)
has to be much greater than the right-handed one ($m_{\tilde{t}_R}$).
According to \cite{Carena:2012np}, $m_{\tilde{q}}$ may be as large as $\mathcal{O}(10^6)$ TeV.
In such a case, the effective theory approach is more appropriate.  
The effective theory of the LSS is constructed in~\cite{Carena:2008rt} 
and is applied to the EWBG~\cite{Carena:2008vj,Carena:2012np}.
Here, to make our analysis simple, we take $m_h$ as an input.

In the limit of $m_{\tilde{q}}\gg m_{\tilde{t}_R}$, the stop masses are reduced to
\begin{align}
\bar{m}_{\tilde{t}_2}^2 &= m_{\tilde{q}}^2
	+\frac{y_t^2\sin^2\beta}{2}\left(1+\frac{|X_t|^2}{m_{\tilde{q}}^2}\right)\varphi^2+\mathcal{O}(g^2)
	\simeq m_{\tilde{q}}^2, \\
\bar{m}_{\tilde{t}_1}^2 &= m_{\tilde{t}_R}^2 
	+\frac{y_t^2\sin^2\beta}{2}\left(1-\frac{|X_t|^2}{m_{\tilde{q}}^2}\right)\varphi^2+\mathcal{O}(g^2),
\end{align}
where $X_t=A_t-\mu/\tan\beta$. 
As discussed above, the heavy stop does not play any role as far as the first-order EWPT is concerned.
Also, the small $m_{\tilde{t}_R}^2$ and $X_t$ are desirable. More precisely, 
at finite temperature, the light stop receives a thermal correction ($\Delta m_{\tilde{t}_R}^2(T)$) 
which is of the order of $T^2$ to leading order.
So the $m_{\tilde{t}_R}^2 +\Delta m_{\tilde{t}_R}^2(T)$ should be vanishingly small
for the strong first-order EWPT, which implies $m_{\tilde{t}_R}^2<0$ since $\Delta m_{\tilde{t}_R}^2(T)>0$.  
In the LSS scenario considered in~\cite{Carena:2008vj}, 
the color-charge-breaking vacuum is the global minimum.
On the other hand, the electroweak vacuum is metastable
whose lifetime is longer than the age of the universe.

For simplicity, we take $m_{\tilde{t}_R}^2=0$ in the following discussion.
The coefficient of cubic term in the one-loop effective potential at finite temperature is
\begin{align}
V_{\rm eff}\ni -(E_{\rm SM}+E_{\tilde{t}_1})T\varphi^3, \quad
E_{\tilde{t}_1} = \frac{y_t^3\sin^3\beta}{4\sqrt{2}}\left(1-\frac{|X_t|^2}{m_{\tilde{q}}^2}\right)^{3/2}.
\end{align}
The deviation of $\lambda_{hhh}^{\rm MSSM}$ from the SM value is found to be~\cite{Kanemura:2004ch}
\begin{align}
\frac{\Delta\lambda_{hhh}^{\rm MSSM}}{\lambda_{hhh}^{\rm SM}}
\simeq \frac{m_t^4}{2\pi^2v^2m_h^2}\left(1-\frac{|X_t|^2}{m_{\tilde{q}}^2}\right)^3
=\frac{2v^4}{m_t^2m_h^2}E_{\tilde{t}_1}^2\simeq 0.05,
\end{align}
where we take $m_h=126$ GeV, $\tan\beta=10$ and $X_t=0$ in the last step.
Unlike the 2HDM, the size of the loop effect is bounded by the top Yukawa coupling.
Beyond the MSSM, however, this limit would be relaxed. 
For instance, in a SUSY model proposed in~\cite{Kanemura:2012uy}, 
the large enhancement is observed~\cite{EWPTvsHHH_SUSY}.
More examples along the same line can be found in \cite{Noble:2007kk}.

\section{Summary}
In this talk, the EWBG is briefly reviewed, focusing on 
the nondecoupling loop properties that may induce the strong first-order EWPT 
as well as the large quantum corrections to the triple Higgs boson coupling.
The 2HDM is the typical example that has the large nondecoupling property. 
Since the deviation of the triple Higgs boson coupling from the SM value is rather large,
its detection at colliders~\cite{HHH_colliders} 
would be one of the experimental indications of the EWBG.

\begin{acknowledgments}
Author would like to thank the organizers of the HPNP 2013 for great hospitality at the workshop.
\end{acknowledgments}

\bigskip 

\begin{thebibliography}{99} 

\bibitem{Beringer:1900zz} 
  J.~Beringer {\it et al.}  [Particle Data Group Collaboration],
  Phys.\ Rev.\ D {\bf 86}, 010001 (2012).

\bibitem{Sakharov:1967dj} 
  A.~D.~Sakharov,
  Pisma Zh.\ Eksp.\ Teor.\ Fiz.\  {\bf 5}, 32 (1967)
  [JETP Lett.\  {\bf 5}, 24 (1967)]
  [Sov.\ Phys.\ Usp.\  {\bf 34}, 392 (1991)]
  [Usp.\ Fiz.\ Nauk {\bf 161}, 61 (1991)].

\bibitem{Shaposhnikov:2009zzb} 
  M.~Shaposhnikov,
  J.\ Phys.\ Conf.\ Ser.\  {\bf 171}, 012005 (2009).

\bibitem{ewbg}
  V.~A.~Kuzmin, V.~A.~Rubakov and M.~E.~Shaposhnikov,
  Phys.\ Lett.\ B {\bf 155} (1985) 36.
For reviews on electroweak baryogenesis, see
A.~G.~Cohen, D.~B.~Kaplan and A.~E.~Nelson,
Ann.\ Rev.\ Nucl.\ Part.\ Sci.\  {\bf 43} (1993) 27;
%
M.~Quiros,
Helv.\ Phys.\ Acta {\bf 67} (1994) 451;
%
V.~A.~Rubakov and M.~E.~Shaposhnikov,
Usp.\ Fiz.\ Nauk {\bf 166} (1996) 493;
%
K.~Funakubo,
Prog.\ Theor.\ Phys.\  {\bf 96} (1996) 475;
%
M.~Trodden,
Rev.\ Mod.\ Phys.\  {\bf 71} (1999) 1463;
%
W.~Bernreuther,
Lect.\ Notes Phys.\  {\bf 591} (2002) 237;
%
  J.~M.~Cline,
  [arXiv:hep-ph/0609145];
%
%
  D.~E.~Morrissey and M.~J.~Ramsey-Musolf,
  New J.\ Phys.\  {\bf 14}, 125003 (2012).
%
  T.~Konstandin,
  arXiv:1302.6713 [hep-ph].

\bibitem{CKM}
  N.~Cabibbo,
  Phys.\ Rev.\ Lett.\  {\bf 10}, 531 (1963);~
  %
  M.~Kobayashi and T.~Maskawa,
  Prog.\ Theor.\ Phys.\  {\bf 49}, 652 (1973).
  
\bibitem{Funakubo:2009eg} 
  K.~Funakubo and E.~Senaha,
  Phys.\ Rev.\ D {\bf 79}, 115024 (2009).

\bibitem{sph_SM}
  N.~S.~Manton,
  Phys.\ Rev.\ D {\bf 28}, 2019 (1983);~
%
  F.~R.~Klinkhamer and N.~S.~Manton,
  Phys.\ Rev.\ D {\bf 30}, 2212 (1984).

\bibitem{sph_wU1Y}
  B.~Kleihaus, J.~Kunz and Y.~Brihaye,
  Phys.\ Lett.\ B {\bf 273}, 100 (1991);~
%
  F.~R.~Klinkhamer and R.~Laterveer,
  Z.\ Phys.\ C {\bf 53}, 247 (1992).
  
\bibitem{Chung:2012vg} 
  D.~J.~H.~Chung, A.~J.~Long and L.~-T.~Wang,
  Phys.\ Rev.\ D {\bf 87}, 023509 (2013).

\bibitem{Funakubo:2012qc} 
  K.~Funakubo and E.~Senaha,
  Phys.\ Rev.\ D {\bf 87}, 054003 (2013).

\bibitem{Patel:2011th} 
  H.~H.~Patel and M.~J.~Ramsey-Musolf,
  JHEP {\bf 1107}, 029 (2011).

\bibitem{Wainwright:2012zn} 
  C.~L.~Wainwright, S.~Profumo and M.~J.~Ramsey-Musolf,
  Phys.\ Rev.\ D {\bf 86}, 083537 (2012).

\bibitem{Garny:2012cg} 
  M.~Garny and T.~Konstandin,
  JHEP {\bf 1207}, 189 (2012).

\bibitem{sm_ewpt}
  K.~Kajantie, M.~Laine, K.~Rummukainen and M.~E.~Shaposhnikov,
  Phys.\ Rev.\ Lett.\  {\bf 77}, 2887 (1996);~
%
  K.~Rummukainen, M.~Tsypin, K.~Kajantie, M.~Laine and M.~E.~Shaposhnikov,
  Nucl.\ Phys.\  B {\bf 532}, 283 (1998);~
%
  F.~Csikor, Z.~Fodor and J.~Heitger,
  Phys.\ Rev.\ Lett.\  {\bf 82}, 21 (1999);~
%
  Y.~Aoki, F.~Csikor, Z.~Fodor and A.~Ukawa,
  Phys.\ Rev.\  D {\bf 60}, 013001 (1999).

\bibitem{Kanemura:2004ch} 
  S.~Kanemura, Y.~Okada and E.~Senaha,
  Phys.\ Lett.\ B {\bf 606}, 361 (2005).

\bibitem{HHH_2HDM}
  S.~Kanemura, S.~Kiyoura, Y.~Okada, E.~Senaha and C.~P.~Yuan,
  Phys.\ Lett.\ B {\bf 558}, 157 (2003);~
%
  S.~Kanemura, Y.~Okada, E.~Senaha and C.~-P.~Yuan,
  Phys.\ Rev.\ D {\bf 70}, 115002 (2004).

\bibitem{LSS}
  M.~S.~Carena, M.~Quiros and C.~E.~M.~Wagner,
  Phys.\ Lett.\ B {\bf 380}, 81 (1996);~
%
  D.~Delepine, J.~M.~Gerard, R.~Gonzalez Felipe and J.~Weyers,
  Phys.\ Lett.\ B {\bf 386}, 183 (1996).

\bibitem{LHCtension}  
  T.~Cohen, D.~E.~Morrissey and A.~Pierce,
  Phys.\ Rev.\ D {\bf 86}, 013009 (2012);~
%
  D.~Curtin, P.~Jaiswal and P.~Meade,
  JHEP {\bf 1208}, 005 (2012);~
 %
  K.~Krizka, A.~Kumar and D.~E.~Morrissey,
  arXiv:1212.4856 [hep-ph].
\bibitem{Carena:2012np} 
  M.~Carena, G.~Nardini, M.~Quiros and C.~E.~M.~Wagner,
  JHEP {\bf 1302}, 001 (2013).
%

\bibitem{Carena:2008rt} 
  M.~Carena, G.~Nardini, M.~Quiros and C.~E.~M.~Wagner,
  JHEP {\bf 0810}, 062 (2008).
%
\bibitem{Carena:2008vj} 
  M.~Carena, G.~Nardini, M.~Quiros and C.~E.~M.~Wagner,
  Nucl.\ Phys.\ B {\bf 812}, 243 (2009).

\bibitem{Kanemura:2012uy} 
  S.~Kanemura, T.~Shindou and T.~Yamada,
  Phys.\ Rev.\ D {\bf 86}, 055023 (2012).

\bibitem{EWPTvsHHH_SUSY}
  S.~Kanemura, E.~Senaha, T.~Shindou and T.~Yamada,
  arXiv:1211.5883 [hep-ph]. See also the talk given by
%
%
  T.~Yamada,
  arXiv:1304.5029 [hep-ph].

\bibitem{Noble:2007kk} 
  A.~Noble and M.~Perelstein,
  Phys.\ Rev.\ D {\bf 78}, 063518 (2008).

 \bibitem{HHH_colliders}
  A.~Djouadi, W.~Kilian, M.~Muhlleitner and P.~M.~Zerwas,
  Eur.\ Phys.\ J.\ C {\bf 10} (1999) 27.
  %
  K. Fujii, talk given at LCWS12,
  ``Higgs Physics at the LC and Requirements"
  http://www.uta.edu/physics/lcws12/.
 %

\bibitem{Funakubo:1993jg} 
  K.~Funakubo, A.~Kakuto and K.~Takenaga,
  Prog.\ Theor.\ Phys.\  {\bf 91}, 341 (1994).

\end{thebibliography}

\end{document}